\documentstyle[epsfig]{aipproc}

\newcommand{\beq}{\begin{equation}}
\newcommand{\eeq}{\end{equation}}
\newcommand{\beqar}{\begin{eqnarray}}
\newcommand{\eeqar}{\end{eqnarray}}

\begin{document}
\title{An Anisotropic Illumination Model\\ of Seyfert I Galaxies}

\author{P.O. Petrucci$^*$, G. Henri$^*$, J. Malzac$^{**}$ and 
E. Jourdain$^{**}$} 
\address{$^*$Laboratoire d'Astrophysique, Grenoble FRANCE\\
$^{\dagger}$Centre d'Etude Spatiale des Rayonnements, (CNRS/UPS), Toulouse FRANCE}

%\lefthead{LEFT head}
%\righthead{RIGHT head}
\maketitle

\begin{abstract}
We develop a self-consistent model of Seyfert galaxies continuum emission.
%High energy emission is produced by Inverse Compton process on soft
%photons emitted as thermal radiation by the accretion disk. Thermal
%emission is, at turn, entirely due to reprocessing of the impinging high
%energy photons.   
The high energy source is assumed to be an optically thin
plasma of highly relativistic leptons ($e^+-e^-$), at rest at a given
height on the disk axis.
% This warm source could be the result of a strong
%schock between an 
%abortive jet coming from the disk and the interstellar medium.
 Such a geometry is highly anisotropic, which has a strong influence on
 Compton process. 
Monte-Carlo simulations allow the superposition of a reflected component
%(the so-called high energy bump)
 to the UV to X-ray spectrum obtained with
our model, leading us to a first comparison with
observations by fitting the high energy spectra of NGC4151 and IC4329a.
\end{abstract}

%%%%%%%%%%%%%%%%%%%%%%%%%%%%%%%%%%%%%%%%%%%%%%%%%%%%%%%%%%%%%%%%%%%%%%%%%%%%%%
\section*{Introduction}
\begin{figure}[t!] % fig 1
\centerline{\epsfig{file=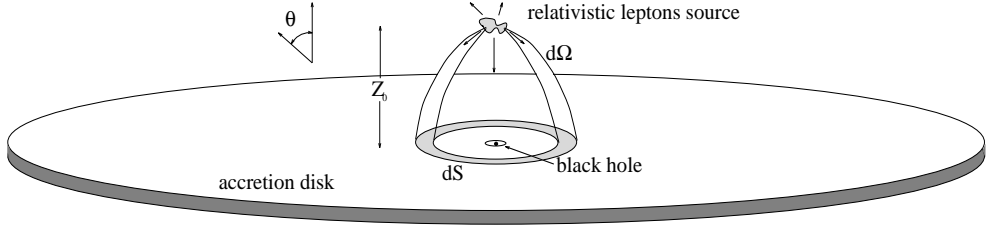,height=13cm,width=3cm,angle=-90}}
\vspace{10pt}
\caption{The general picture of the model. We have also drawn the
    trajectory of a beam of photons  emitted by the hot source in a solid
    angle $d\Omega$ and absorbed by a surface ring $dS$ on the
    disk.}
\label{main}
\vspace{-0.5cm}
\end{figure}

%%%%%%%%%%%%%%%%%%%%%%%%%%%%%%%%%%%%%%%%%%%%%%%%%%%%%%%%%%%%%%%%%%%%%%%%%%%%%%
%It is widely believed that the high energy emission of AGNs is produced
%by Comptonization of soft photons by high energy electrons or pairs.
%Besides, for Seyfert galaxies, some observational facts support 
%the idea that high energy radiation can be primarily produced and 
%reflected on a cold surface, producing a fair fraction of thermal 
%UV-optical radiation (\cite*{Cla92},\cite*{Pou90}). 
We propose a new
model, for Seyfert galaxies, involving a point source of relativistic
leptons located above the disk (that could be physically realized by a
strong shock terminating an aborted jet)
emitting hard radiation by Inverse Compton (IC) process on soft photons
produced by the accretion disk. The disk itself radiates only through the
re-processing of the hard radiation impinging on it, i.e. we do not suppose
any internal energy dissipation (cf Fig. \ref{main}).
%(this is a relatively good approximation
%since the disk supplies its jet with almost all the available gravitational
%power, being then weakly dissipative (\cite*{Ferr})) .
 Such a geometry is
highly anisotropic, which takes a real importance in the computation of IC
process (\cite{Ghi91}, \cite{Hen*}). We treat both Newtonian and general
relativistic cases \cite{Pet*}, deriving a self-consistent solution in the Newtonian
case. We have recently added Monte-Carlo calculations 
%developped by J. Malzac and E. Jourdain 
to take into account, in the high energy part of our synthetic spectra,
the compton reflection component and the fluorescent iron line \cite{Pou90}.
\noindent
We present here
%the main equations describing the radiative balance between the hot
%source and the disk, as well as 
the most important results supplied by
the model. 

\begin{figure}[b!] % fig 1
\centerline{\epsfig{file=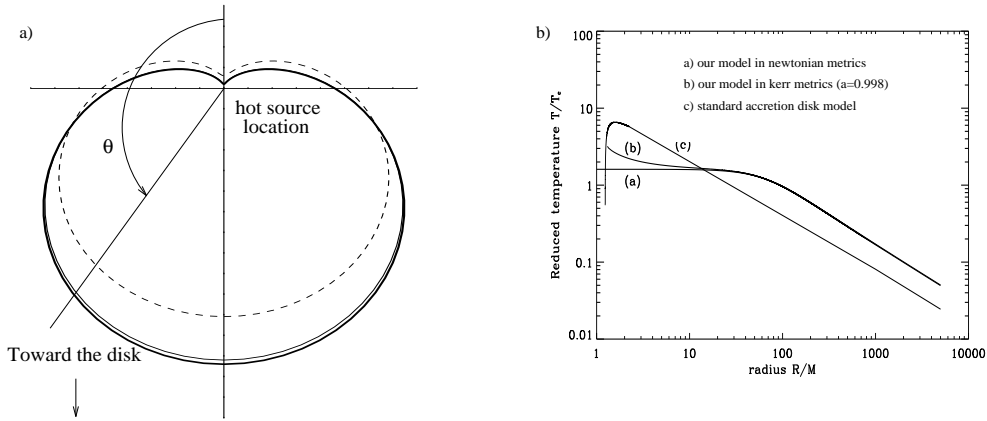,height=13cm,width=5.5cm,angle=-90}}
\vspace{10pt}
\caption{(a) Polar plots of $\displaystyle\frac{dP}{d\Omega}$ for
     $Z_0/M=100$ (solid line) and $Z_0/M=10$ (dashed line) in Kerr
     metrics with $a=0.998$. The bold line corresponds to the Newtonian
     metrics. (b) Effective temperature versus r for $Z_0/M=70$}
\label{anisotropytemperature}
\end{figure}
%%%%%%%%%%%%%%%%%%%%%%%%%%%%%%%%%%%%%%%%%%%%%%%%%%%%%%%%%%%%%%%%%%%%%%%%%%%%%%
\section*{Angular distribution of the hot source}
%%%%%%%%%%%%%%%%%%%%%%%%%%%%%%%%%%%%%%%%%%%%%%%%%%%%%%%%%%%%%%%%%%%%%%%%%%%%%%
It appears that the anisotropy of the soft photon
field at the hot source level, leads to an {\bf anisotropic Inverse
Compton process}, with much more radiation being scattered backward than
forward. Such an
anisotropic re-illumination could naturally explain the apparent X-ray
luminosity, usually much lower than the optical-UV continuum emitted in
the blue bump \cite{WalF93}. It can also explain the equivalent width
observed for the 
iron line, which requires more impinging radiation than what is actually
observed \cite{Nan97}. 
 We plot in
Figure \ref{anisotropytemperature} the angular distribution of the power
emitted by the hot source in Newtonian metrics and for different values
of the source height in Kerr metrics. It appears that 
the closer the source to the black hole is, the less anisotropic the high
energy photon field is. This is principally due to the curvature of geodesics
making the photons emitted near the black hole arrive at larger angle
than in the Newtonian case.

%%%%%%%%%%%%%%%%%%%%%%%%%%%%%%%%%%%%%%%%%%%%%%%%%%%%%%%%%%%%%%%%%%%%%%%%%%%%%%
\section*{Disk temperature profile}
%%%%%%%%%%%%%%%%%%%%%%%%%%%%%%%%%%%%%%%%%%%%%%%%%%%%%%%%%%%%%%%%%%%%%%%%%%%%%%
The radiative balance between the hot source and the disk allows to
compute the temperature profile on the disk surface. It is, in
fact, markedly different from ``standard accretion disk model'' as shown
in Figure \ref{anisotropytemperature}. Indeed, even if at large distances, all
models give the same asymptotic behavior $T\propto R^{-3/4}$, in the
inner part of the disk, it keeps increasing in ``standard model''
whereas, in our model, for $R\leq Z_0$, the 
{\bf temperature saturates} around a characteristic value $T_c$.
%\begin{figure}[b!] % fig 1
%\centerline{\epsfig{file=temperature.ps,height=6cm,width=8cm}}
%\vspace{10pt}
%\caption{Effective temperature versus r for $Z_0/M=70$\\
%                            a) Our model in Newtonian metrics\\
%                            b) Our model in Kerr metrics\\
%                            c) Standard accretion disk 
% }\label{temperature}
%\end{figure}
%Indeed, the power radiated by the disk is essentially controlled by the
%angular distribution of the hot source $\displaystyle\frac{dP}{d\Omega}$
%(cf Eq. \ref{eqF}) which is approximatively constant for $R\leq Z_0$
%(i.e. $\theta\simeq\pi /4$). 
The differences between Newtonian and Kerr
metrics come only from Gravitational and Doppler shifts, which are only
appreciable for $R\leq 5M$. 
%Thus, unless $Z_0$ is itself small enough,
%these shifts concern only a small fraction of the emitting area at
%$T=T_c$, and modified hardly the UV to X-ray spectrum.

%%%%%%%%%%%%%%%%%%%%%%%%%%%%%%%%%%%%%%%%%%%%%%%%%%%%%%%%%%%%%%%%%%%%%%%%%%%%%%
%\section*{The overall spectra}
%%%%%%%%%%%%%%%%%%%%%%%%%%%%%%%%%%%%%%%%%%%%%%%%%%%%%%%%%%%%%%%%%%%%%%%%%%%%%%
%The overall UV to X-ray spectra can be deduced from this model. The bulk of
%the energy coming from the disk is emitted on the blue and the
%ultraviolet, giving the well-known ``blue-bump'' observed in most quasars
%and many AGNs. On the other hand, the high energy spectrum is a power law
%with a exponentially cut-off. It depends directly on the relativistic
%particle distribution adopted. The shape of the spectra are highly
%dependent on the inclination angle and, for the relativistic cases, on the
%height of the hot source above the disk $Z_0$.
\begin{figure}[t!] % fig 1
\centerline{\epsfig{file=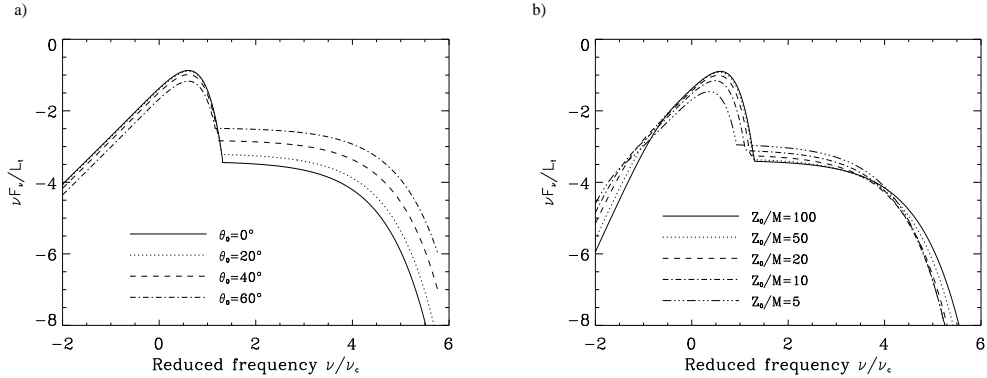,height=13cm,width=5cm,angle=-90}}
\vspace{10pt}
\caption{(a) Differential power spectrum for different values
          of the inclination angle in the Newtonian case. (b)
          Differential power spectrum for different values 
          of $Z_0$ for the Kerr maximal case. We use reduced
          coordinates.}
\label{specangzovar} 
\end{figure}

%%%%%%%%%%%%%%%%%%%%%%%%%%%%%%%%%%%%%%%%%%%%%%%%%%%%%%%%%%%%%%%%%%%%%%%%%%%%%%
\section*{Influence of the inclination angle}
%%%%%%%%%%%%%%%%%%%%%%%%%%%%%%%%%%%%%%%%%%%%%%%%%%%%%%%%%%%%%%%%%%%%%%%%%%%%%%
The overall UV to X-ray spectra can be deduced from this model. The bulk of
the energy coming from the disk gives the well-known ``blue-bump''.
 On the other hand, in order to avoid run-away electrons, the relativistic
particle distribution is supposed to be a {\bf power law
with an exponential cut-off} about 300 keV, giving  the high energy part
of the spectrum. One can see on Figures \ref{specangzovar}a Newtonian spectra
for different inclination angles for $Z_0/M=10$. For all
inclination angles, the Kerr spectra are always weaker in UV and
 brighter in X-ray than the Newtonian ones.
% However, the difference tends
%to be less visible for the highest inclination angles. This can be
%explained easily by the fact that, for high inclination angle, the part
%of the disk moving toward the 
%observer emits blue-shifted radiation, compensated by the red-shifted
%radiation from the other parts.
These effects are much less pronounced
for high $Z_0/M$ values because the emission area is much larger, and
thus is less affected by relativistic corrections.  

%%%%%%%%%%%%%%%%%%%%%%%%%%%%%%%%%%%%%%%%%%%%%%%%%%%%%%%%%%%%%%%%%%%%%%%%%%%%%%
\section*{Influence of the hot source height}
%%%%%%%%%%%%%%%%%%%%%%%%%%%%%%%%%%%%%%%%%%%%%%%%%%%%%%%%%%%%%%%%%%%%%%%%%%%%%%
Figure \ref{specangzovar}b shows the overall spectrum, for
different values of $Z_0/M$ in Kerr metrics for $\theta=0^\circ$. The
relativistic effects become important for values of 
$Z_0/M$ smaller than about $50$. They produce a variation of intensity
lowering the blue-bump and increasing the hard X-ray emission. The change
in the UV range is due to the transverse Doppler effect between the rotating
disk and the observer, producing a net red-shift. In the X-ray range,
the variation is due to the high energy dependence on $Z_0/M$ (cf Figure
\ref{anisotropytemperature}). The observed {\bf X/UV 
ratio} can then be strongly altered by these effects. Quantitatively, the 
luminosity ratio between the maximum of the blue-bump and the X-ray
plateau goes from $\simeq 300$ in the Newtonian case, to $\simeq 10$ for 
$Z_0/M=3$.

%%%%%%%%%%%%%%%%%%%%%%%%%%%%%%%%%%%%%%%%%%%%%%%%%%%%%%%%%%%%%%%%%%%%%%%%%%%%%%
\section*{Scaling laws}
%%%%%%%%%%%%%%%%%%%%%%%%%%%%%%%%%%%%%%%%%%%%%%%%%%%%%%%%%%%%%%%%%%%%%%%%%%%%%%
\begin{figure}[b!] % fig 1
\centerline{\epsfig{file=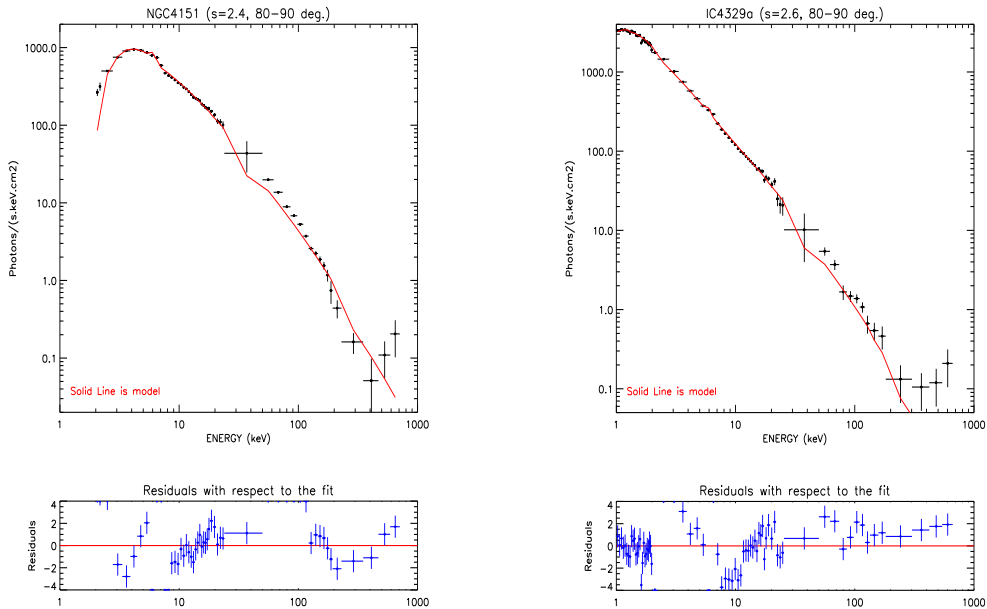,height=13cm,width=8cm,angle=-90}}
\vspace{10pt}
%\caption{fits of IC4329a and NGC4151}
\caption{The data are deconvolued
  ROSAT/GINGA/OSSE data published respectively in 
%\cite{Zdz96}, and 
% \cite{Mad95}
 [4] and [11]. Note that in both spectra we added a fictive data
 point around 40 kev with large error bars to join GINGA and OSSE data.    
% The model is our high energy primary spectrum, plus the reflection
% component,  
% plus a simple neutral absorber. 
 The parameters are the electron index
 $s$, the high energy cut-off $\nu_0$, the viewing angle $\theta$ and the
 absorber column  density $N_h$. In these fits the only free parameter is
 $N_h$, while the others are  kept at a convenient value. For both fits
 $\nu_0=100kev$, $\theta$=80-90 deg. The best fit value for $N_h$ is 
 $1.2 10^{23}$ for NGC4151 and $4.6 10^{21}$ for IC4329a.
 The reduced $\chi^2$ are respectively 12.0 and 5.92}
\label{reflection} 
\end{figure}
With some further assumptions, the model predicts scaling laws quite different
from the standard accretion models. If one assumes a constant high energy
cut-off (possibly fixed by the pair production threshold) and a
constant solid angle subtended by the hot source, then the 
  following mass scaling laws apply:
\begin{eqnarray}
        T_{c} & = & constant\nonumber  \\
        L_{c} & \propto & M^{2} \nonumber
\end{eqnarray}
The first equality could explain the weak variations of the blue bump
component, even though the luminosity L ranges over 6 orders of magnitude
from source to source \cite{WalF93}. The second one is in agreement
with the results of \cite{Col91} where they find in a sample of Seyfert
I and quasars a correlation between the mass and the luminosity under the
form $L\propto M^{\beta}$  with $\beta=1.8\pm 0.6$.

%%%%%%%%%%%%%%%%%%%%%%%%%%%%%%%%%%%%%%%%%%%%%%%%%%%%%%%%%%%%%%%%%%%%%%%%%%%%%%
\section*{The reflection component}
%%%%%%%%%%%%%%%%%%%%%%%%%%%%%%%%%%%%%%%%%%%%%%%%%%%%%%%%%%%%%%%%%%%%%%%%%%%%%%
We considered an infinite disc of neutral matter with solar abondances (
opacities given by \cite{Mor83}), and hydrogen column density of
$10^{25}cm^{-2}$. We found that about 10 \% of the total energy emitted
toward the disc is reflected and does not contribute to the disc heating.
However, the reflection component has important effects on the shape of the  
observed spectrum. Due to the strong anisotropy of the primary source,
the ratio of the reflected to the primary component can be very high. The
spectrum strongly depends on the viewing angle $\theta$ with respect to
the disc normal. For nearly edge on inclinations, the reflection
contribution is weak. As the inclination decrease the reflection
component increase \cite{Mag95}. Conversely, the primary
spectrum decreases with decreasing $\theta$. So that, for a face on
inclination, reflection dominates the hard X-ray spectrum. A rough
comparison to IC4329a and NGC 4151 spectra shows that 
our model is reasonnably close to the data if the orientation is nearly
edge-on (cf. fig \ref{reflection}).
 We can note that the anisotropic illumination model
predicts a weak UV component for high inclination angles which is in
qualitative agreement with the observed UV to soft X-ray luminosity ratio
in these two objects.\\
% We are currently working to get more rigorous
%comparisons to a larger sample of Seyfert galaxies. \\
\\
{\bf Acknowledgment :} We are very grateful to A. Zdziarski for providing
us the NGC4151 and IC4329a data.

%%%%%%%%%%%%%%%%%%%%%%%%%%%%%%%%%%%%%%%%%%%%%%%%%%%%%%%%%%%%%%%%%%%%%%%%%%%%%%%
%\section*{CONCLUSIONS}
%%%%%%%%%%%%%%%%%%%%%%%%%%%%%%%%%%%%%%%%%%%%%%%%%%%%%%%%%%%%%%%%%%%%%%%%%%%%%%%
%\label{conclusionsection}
%We have shown that a model based on reillumination of a disk by an 
%anisotropic IC source could lead to a self-consistent picture 
%where the angular distribution of high energy radiation and the 
%radial temperature distribution of the disk are mutually linked and 
%both  
%determined in a single way. The model offers a simple explanation 
%for the correlated long term variability of X and UV radiation, 
%the short term variability of X-rays non correlated with UV 
%variations, and the apparent X/UV deficit that seems 
%contradictory with simple reillumination models. In its simplest 
%form, it predicts a unique shape of disk spectrum and a X/UV ratio 
%depending only on the inclination angle. The predicted values are in 
%good agreement with observations. A precise comparison with real 
%spectra should also include other components, such as a 
%reflection component and a fluorescent Fe K$\alpha$ line. This is 
%deferred to a future work. Finally, a more complete work has to be done
%to explain the exact mechanism of emission of the hot source, supposed to
%be realized by a strong shock terminating an aborted jet.
%

\end{document}